\documentclass[acmsmall,screen,nonacm]{acmart}

\usepackage{xspace}
\usepackage{colortbl}
\usepackage{array}

\definecolor{lightgreen}{rgb}{0.8,1,0.8}
\definecolor{lightred}{rgb}{1,0.8,0.8}
\definecolor{darkgreen}{rgb}{0.4,1,0.4}
\definecolor{orange}{rgb}{1,0.9,0.7}

\AtBeginDocument{%
  }

\renewcommand\footnotetextcopyrightpermission[1]{} 
\renewcommand\footnotetextcopyrightpermission[1]{} 

\renewcommand\footnotetextcopyrightpermission[1]{} 

\acmConference[ICTD '24]{Information and Communication Technologies and Development}{Dec 09--11, 2020}{Nairobi, Kenya}

\def\CI{C\^{o}te d'Ivoire\xspace}
\def\Meagui{\textit{RegionM}\xspace}

\def\TaRL{\textit{newMethod}\xspace}
\def\chatbot{conversational agent }

\begin{document}

\title{Applying Think-Aloud in ICTD: A Case Study \\ of a Chatbot Use   by Teachers in Rural \CI}


\author{Vikram Kamath Cannanure}
\orcid{0000-0002-0944-7074}
\email{vica001@teams.uni-saarland.de}
\affiliation{%
  \institution{Saarland University}
  \city{Saarbrücken}
  \country{Germany}
}

\author{Sharon Wolf}
\orcid{0000-0002-8076-8399}
\email{wolfs@upenn.edu}
\affiliation{%
  \institution{University of Pennsylvania}
  \city{Philadelphia}
  \country{USA}
}

\author{Kaja Jasińska}
\orcid{0000-0002-8851-1627}
\email{kaja.jasinska@utoronto.ca}
\affiliation{%
  \institution{University of Toronto}
  \city{Toronto}
  \country{Canada}
}
\author{Timothy X Brown}
\orcid{0000-0002-0670-0935}
\email{timxb@cmu.edu}
\affiliation{%
 \institution{Carnegie Mellon University Africa}
  \city{Kigali}
  \country{Rwanda}
}

\author{Amy Ogan}
\orcid{0000-0003-2671-6149}
\email{aeo@cs.cmu.edu}
\affiliation{%
  \institution{Carnegie Mellon University}
  \city{Pittsburgh}
  \country{USA}
}

\renewcommand{\shortauthors}{Cannanure et al.}

\begin{abstract}
Think-alouds are a common HCI usability method where participants verbalize their thoughts while using interfaces. However, their utility in cross-cultural settings, particularly in the Global South, is unclear, where cultural differences impact user interactions. This paper investigates the usability challenges teachers in rural \CI faced when using a chatbot designed to support an educational program. We conducted think-aloud sessions with 20 teachers two weeks after a chatbot deployment, analyzing their navigation, errors, and time spent on tasks. We discuss our approach and findings that helped us identify usability issues and challenging features for improving the chatbot designs. Our note summarizes our reflections on using think-aloud and contributes to discussions on its culturally sensitive adaptation in the Global South.
\end{abstract}

\ccsdesc[500]{Human-centered computing~Empirical studies in HCI}
\ccsdesc[300]{Applied computing~Interactive learning environments}
\ccsdesc[100]{Social and professional topics~Cultural characteristics}

\keywords{Think-Aloud,  Cross-cultural usability, Chatbot, Design, HCI4D, ICTD}

\maketitle

\section{Introduction}


Think-aloud methods are widely used in usability research to gain insights into users’ cognitive processes during task performance \cite{ramey_does_2006,nielsen_getting_2002}. This method involves participants verbalizing their thoughts while completing a task, providing researchers access to unobservable mental processes during user interactions \cite{charters_use_2003}. Think-aloud has been applied in various fields in HCI, such as education and qualitative research \cite{garcia_uso_2018}.  Proper design, test procedures, and analysis are integral to using this method \cite{nielsen_getting_2002}.  However, while the think-aloud method offers valuable insights in Western contexts, researchers must remain aware of potential limitations and explore opportunities to adapt this method to different contexts appropriately \cite{nielsen_getting_2002}.

Applying think-aloud techniques in the Global South introduces additional challenges due to cultural differences. Think-Aloud research across Africa, India, and Denmark has shown that traditional Western usability testing methods are less utility in cross-cultural settings \cite{clemmensen_cross_2006,oyugi_evaluation_2008}, highlighting variations in personal interaction norms \cite{oyugi_evaluation_2008}. Recognizing and addressing these cultural factors is vital when conducting user-centered design testing in diverse contexts \cite{perri_crucial_2014,dell_ins_2016}. The utility of think-aloud tests is often maximized when evaluators and test users share the same cultural background \cite{clemmensen_cross_2006}, thus requiring the capacity building to train prospective researchers. As HCI4D grows, it is increasingly important to adapt methodologies for cultural relevance to serve under-resourced populations better worldwide \cite{dell_ins_2016}.

Our project describes a think-aloud session for teachers using a chatbot in rural \CI. Recently, chatbots have become popular in HCI, providing natural language interfaces to digital systems \cite{brandtzaeg_why_2017,DIA,Jain2018a}. Think-aloud techniques remain a prominent method to understand chatbot usability and design \cite{nielsen_getting_2002,barbosa_ux_2022}. \citet{barbosa_ux_2022} found that think-aloud was a favorable method for users using a chatbot, and it helped capture users' insights over other methods \cite{barbosa_ux_2022}. As chatbots evolve, further refinement of UX evaluation methods is needed to address new challenges \cite{hristidis_chatbot_2018}.

\begin{figure}[!htp]
\centering
\includegraphics[width=\textwidth]{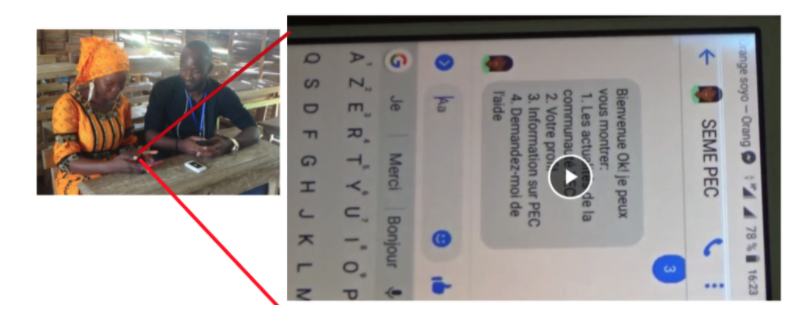}
\caption{Think-aloud session with a teacher, conducted two weeks after training. The teachers were asked to use the chatbot features as the interviewer recorded their attempts. They were given three attempts per feature, and the entire session was recorded on video. }
\label{think-aloud-sessions}
\end{figure}

We conducted 20 think-aloud sessions (see Figure~\ref{think-aloud-sessions}) with rural teachers to observe how they interact with the chatbot. Teachers were asked to navigate the chatbot's different features (see Table~\ref{tab:stats}) two weeks after deployment. Through our work, we hope to contribute our reflections and discussions for adapting think-aloud to the Global South.


\section{Methodology}
The study occurred in a southwestern region of \CI in May 2021, where French is the official language, though nearly 70 local languages are spoken~\cite{ethnologue2017}. The main occupation is agricultural, with cocoa and coffee being key income sources for decades~\cite{kiewisch2015looking}. The specific site, \Meagui in the Soubré region, is a rural town with most schools located in remote areas facing significant infrastructure challenges, such as limited access to water, electricity, and phone signals, and poor road conditions, which contribute to low literacy rates among students~\cite{madaio2019everyone,madaio2019you}. The study focused on teachers implementing the \TaRL (anonymized) program to improve basic math and French skills among 3rd, 4th, and 5th graders in rural schools. \TaRL program groups students by proficiency level, and teachers conduct 45-minute daily sessions of interactive, child-centered activities to enhance learning. After gaining initial success, the Ivorian government was interested in scaling up the program with technology~\cite{Study1,Study2}

\subsubsection*{Think-Alouds Sessions}

Two weeks after the chatbot was introduced, a series of think-aloud sessions were conducted with 20 teachers to evaluate their ability to use the different chatbot features (See Appendix \ref{system}). Each teacher was given three attempts to access the feature while verbalizing their thought process. An interviewer observed and recorded their attempts using a survey. For the interviewer, each feature was a survey question with instructions on conducting the think-aloud session, i.e., noting steps taken, success or failure, errors, and time spent. The entire session was video-recorded for further analysis. We also had a follow-up interview with the participants.

\subsubsection*{Data Analysis}
We consolidated field observation notes and interview transcripts into a single dataset. The interview data was transcribed and translated into English, and we applied thematic analysis~\cite{charmaz2008grounded, strauss1990basics, muller2014curiosity, Beyer1998} to identify low-level themes. These themes were synthesized through discussions with the research team and field team to address our research questions.


\section{Findings}


\begin{table}[]
\centering
\caption{The success rate of the think-aloud by feature and teacher. Each row represents a teacher, and each column represents a feature. A green box indicates success and a red box indicates failure. The columns at the end (User Score) indicate the aggregate success percentage for the user. The last row (Feature \%) indicates the aggregate success percentage for the feature. Please see the Appendix \ref{system} for more information}
\label{tab:stats}
\resizebox{\columnwidth}{!}{%
\begin{tabular}{|r|r|r|r|r|r|r|r|r|r|r|r|r|}
\hline
\multicolumn{1}{|l|}{\textbf{Participant}} &
  \multicolumn{1}{l|}{\textbf{Bonjour}} &
  \multicolumn{1}{l|}{\textbf{Survey}} &
  \multicolumn{1}{l|}{\textbf{Goals}} &
  \multicolumn{1}{l|}{\textbf{Browse}} &
  \multicolumn{1}{l|}{\textbf{Tips}} &
  \multicolumn{1}{l|}{\textbf{Manual}} &
  \multicolumn{1}{l|}{\textbf{Ask question}} &
  \multicolumn{1}{l|}{\textbf{Access story}} &
  \multicolumn{1}{l|}{\textbf{Access joke}} &
  \multicolumn{1}{l|}{\textbf{Add story}} &
  \multicolumn{1}{l|}{\textbf{Answer question}} &
  \multicolumn{1}{l|}{\textbf{User score (\%)}} \\ \hline
P1 &
  \cellcolor[HTML]{B7E1CD}1 &
  \cellcolor[HTML]{B7E1CD}1 &
  \cellcolor[HTML]{F4CCCC}0 &
  \cellcolor[HTML]{F4CCCC}0 &
  \cellcolor[HTML]{F4CCCC}0 &
  \cellcolor[HTML]{F4CCCC}0 &
  \cellcolor[HTML]{F4CCCC}0 &
  \cellcolor[HTML]{F4CCCC}0 &
  \cellcolor[HTML]{F4CCCC}0 &
  \cellcolor[HTML]{F4CCCC}0 &
  \cellcolor[HTML]{F4CCCC}0 &
  \cellcolor[HTML]{FFFFFF}\textbf{18.2} \\ \hline
P2 &
  \cellcolor[HTML]{B7E1CD}1 &
  \cellcolor[HTML]{B7E1CD}1 &
  \cellcolor[HTML]{F4CCCC}0 &
  \cellcolor[HTML]{B7E1CD}1 &
  \cellcolor[HTML]{F4CCCC}0 &
  \cellcolor[HTML]{F4CCCC}0 &
  \cellcolor[HTML]{B7E1CD}1 &
  \cellcolor[HTML]{F4CCCC}0 &
  \cellcolor[HTML]{F4CCCC}0 &
  \cellcolor[HTML]{F4CCCC}0 &
  \cellcolor[HTML]{F4CCCC}0 &
  \cellcolor[HTML]{FBDCBC}\textbf{36.4} \\ \hline
P3 &
  \cellcolor[HTML]{B7E1CD}1 &
  \cellcolor[HTML]{B7E1CD}1 &
  \cellcolor[HTML]{F4CCCC}0 &
  \cellcolor[HTML]{F4CCCC}0 &
  \cellcolor[HTML]{B7E1CD}1 &
  \cellcolor[HTML]{B7E1CD}1 &
  \cellcolor[HTML]{F4CCCC}0 &
  \cellcolor[HTML]{F4CCCC}0 &
  \cellcolor[HTML]{F4CCCC}0 &
  \cellcolor[HTML]{F4CCCC}0 &
  \cellcolor[HTML]{F4CCCC}0 &
  \cellcolor[HTML]{FBDCBC}\textbf{36.4} \\ \hline
P4 &
  \cellcolor[HTML]{B7E1CD}1 &
  \cellcolor[HTML]{B7E1CD}1 &
  \cellcolor[HTML]{F4CCCC}0 &
  \cellcolor[HTML]{F4CCCC}0 &
  \cellcolor[HTML]{B7E1CD}1 &
  \cellcolor[HTML]{B7E1CD}1 &
  \cellcolor[HTML]{F4CCCC}0 &
  \cellcolor[HTML]{B7E1CD}1 &
  \cellcolor[HTML]{F4CCCC}0 &
  \cellcolor[HTML]{B7E1CD}1 &
  \cellcolor[HTML]{F4CCCC}0 &
  \cellcolor[HTML]{DEC999}\textbf{54.5} \\ \hline
P5 &
  \cellcolor[HTML]{B7E1CD}1 &
  \cellcolor[HTML]{B7E1CD}1 &
  \cellcolor[HTML]{B7E1CD}1 &
  \cellcolor[HTML]{B7E1CD}1 &
  \cellcolor[HTML]{B7E1CD}1 &
  \cellcolor[HTML]{F4CCCC}0 &
  \cellcolor[HTML]{B7E1CD}1 &
  \cellcolor[HTML]{F4CCCC}0 &
  \cellcolor[HTML]{F4CCCC}0 &
  \cellcolor[HTML]{F4CCCC}0 &
  \cellcolor[HTML]{F4CCCC}0 &
  \cellcolor[HTML]{DEC999}\textbf{54.5} \\ \hline
P6 &
  \cellcolor[HTML]{B7E1CD}1 &
  \cellcolor[HTML]{B7E1CD}1 &
  \cellcolor[HTML]{F4CCCC}0 &
  \cellcolor[HTML]{B7E1CD}1 &
  \cellcolor[HTML]{B7E1CD}1 &
  \cellcolor[HTML]{B7E1CD}1 &
  \cellcolor[HTML]{B7E1CD}1 &
  \cellcolor[HTML]{F4CCCC}0 &
  \cellcolor[HTML]{B7E1CD}1 &
  \cellcolor[HTML]{B7E1CD}1 &
  \cellcolor[HTML]{F4CCCC}0 &
  \cellcolor[HTML]{A8C393}\textbf{72.7} \\ \hline
P7 &
  \cellcolor[HTML]{B7E1CD}1 &
  \cellcolor[HTML]{B7E1CD}1 &
  \cellcolor[HTML]{F4CCCC}0 &
  \cellcolor[HTML]{B7E1CD}1 &
  \cellcolor[HTML]{F4CCCC}0 &
  \cellcolor[HTML]{B7E1CD}1 &
  \cellcolor[HTML]{B7E1CD}1 &
  \cellcolor[HTML]{F4CCCC}0 &
  \cellcolor[HTML]{B7E1CD}1 &
  \cellcolor[HTML]{B7E1CD}1 &
  \cellcolor[HTML]{B7E1CD}1 &
  \cellcolor[HTML]{A8C393}\textbf{72.7} \\ \hline
P8 &
  \cellcolor[HTML]{B7E1CD}1 &
  \cellcolor[HTML]{B7E1CD}1 &
  \cellcolor[HTML]{F4CCCC}0 &
  \cellcolor[HTML]{B7E1CD}1 &
  \cellcolor[HTML]{F4CCCC}0 &
  \cellcolor[HTML]{B7E1CD}1 &
  \cellcolor[HTML]{B7E1CD}1 &
  \cellcolor[HTML]{B7E1CD}1 &
  \cellcolor[HTML]{B7E1CD}1 &
  \cellcolor[HTML]{B7E1CD}1 &
  \cellcolor[HTML]{B7E1CD}1 &
  \cellcolor[HTML]{8DC190}\textbf{81.8} \\ \hline
P9 &
  \cellcolor[HTML]{B7E1CD}1 &
  \cellcolor[HTML]{B7E1CD}1 &
  \cellcolor[HTML]{F4CCCC}0 &
  \cellcolor[HTML]{B7E1CD}1 &
  \cellcolor[HTML]{B7E1CD}1 &
  \cellcolor[HTML]{B7E1CD}1 &
  \cellcolor[HTML]{B7E1CD}1 &
  \cellcolor[HTML]{B7E1CD}1 &
  \cellcolor[HTML]{B7E1CD}1 &
  \cellcolor[HTML]{B7E1CD}1 &
  \cellcolor[HTML]{F4CCCC}0 &
  \cellcolor[HTML]{8DC190}\textbf{81.8} \\ \hline
P10 &
  \cellcolor[HTML]{B7E1CD}1 &
  \cellcolor[HTML]{B7E1CD}1 &
  \cellcolor[HTML]{F4CCCC}0 &
  \cellcolor[HTML]{B7E1CD}1 &
  \cellcolor[HTML]{B7E1CD}1 &
  \cellcolor[HTML]{B7E1CD}1 &
  \cellcolor[HTML]{B7E1CD}1 &
  \cellcolor[HTML]{B7E1CD}1 &
  \cellcolor[HTML]{B7E1CD}1 &
  \cellcolor[HTML]{B7E1CD}1 &
  \cellcolor[HTML]{F4CCCC}0 &
  \cellcolor[HTML]{8DC190}\textbf{81.8} \\ \hline
P11 &
  \cellcolor[HTML]{B7E1CD}1 &
  \cellcolor[HTML]{B7E1CD}1 &
  \cellcolor[HTML]{F4CCCC}0 &
  \cellcolor[HTML]{B7E1CD}1 &
  \cellcolor[HTML]{B7E1CD}1 &
  \cellcolor[HTML]{B7E1CD}1 &
  \cellcolor[HTML]{F4CCCC}0 &
  \cellcolor[HTML]{B7E1CD}1 &
  \cellcolor[HTML]{B7E1CD}1 &
  \cellcolor[HTML]{B7E1CD}1 &
  \cellcolor[HTML]{B7E1CD}1 &
  \cellcolor[HTML]{8DC190}\textbf{81.8} \\ \hline
P12 &
  \cellcolor[HTML]{B7E1CD}1 &
  \cellcolor[HTML]{B7E1CD}1 &
  \cellcolor[HTML]{B7E1CD}1 &
  \cellcolor[HTML]{B7E1CD}1 &
  \cellcolor[HTML]{B7E1CD}1 &
  \cellcolor[HTML]{B7E1CD}1 &
  \cellcolor[HTML]{F4CCCC}0 &
  \cellcolor[HTML]{B7E1CD}1 &
  \cellcolor[HTML]{B7E1CD}1 &
  \cellcolor[HTML]{F4CCCC}0 &
  \cellcolor[HTML]{B7E1CD}1 &
  \cellcolor[HTML]{8DC190}\textbf{81.8} \\ \hline
P13 &
  \cellcolor[HTML]{B7E1CD}1 &
  \cellcolor[HTML]{B7E1CD}1 &
  \cellcolor[HTML]{F4CCCC}0 &
  \cellcolor[HTML]{F4CCCC}0 &
  \cellcolor[HTML]{B7E1CD}1 &
  \cellcolor[HTML]{B7E1CD}1 &
  \cellcolor[HTML]{B7E1CD}1 &
  \cellcolor[HTML]{B7E1CD}1 &
  \cellcolor[HTML]{B7E1CD}1 &
  \cellcolor[HTML]{B7E1CD}1 &
  \cellcolor[HTML]{B7E1CD}1 &
  \cellcolor[HTML]{8DC190}\textbf{81.8} \\ \hline
P14 &
  \cellcolor[HTML]{B7E1CD}1 &
  \cellcolor[HTML]{B7E1CD}1 &
  \cellcolor[HTML]{F4CCCC}0 &
  \cellcolor[HTML]{B7E1CD}1 &
  \cellcolor[HTML]{B7E1CD}1 &
  \cellcolor[HTML]{B7E1CD}1 &
  \cellcolor[HTML]{B7E1CD}1 &
  \cellcolor[HTML]{B7E1CD}1 &
  \cellcolor[HTML]{B7E1CD}1 &
  \cellcolor[HTML]{B7E1CD}1 &
  \cellcolor[HTML]{B7E1CD}1 &
  \cellcolor[HTML]{72BE8D}\textbf{90.9} \\ \hline
P15 &
  \cellcolor[HTML]{B7E1CD}1 &
  \cellcolor[HTML]{B7E1CD}1 &
  \cellcolor[HTML]{B7E1CD}1 &
  \cellcolor[HTML]{B7E1CD}1 &
  \cellcolor[HTML]{B7E1CD}1 &
  \cellcolor[HTML]{F4CCCC}0 &
  \cellcolor[HTML]{B7E1CD}1 &
  \cellcolor[HTML]{B7E1CD}1 &
  \cellcolor[HTML]{B7E1CD}1 &
  \cellcolor[HTML]{B7E1CD}1 &
  \cellcolor[HTML]{B7E1CD}1 &
  \cellcolor[HTML]{72BE8D}\textbf{90.9} \\ \hline
P16 &
  \cellcolor[HTML]{B7E1CD}1 &
  \cellcolor[HTML]{B7E1CD}1 &
  \cellcolor[HTML]{F4CCCC}0 &
  \cellcolor[HTML]{B7E1CD}1 &
  \cellcolor[HTML]{B7E1CD}1 &
  \cellcolor[HTML]{B7E1CD}1 &
  \cellcolor[HTML]{B7E1CD}1 &
  \cellcolor[HTML]{B7E1CD}1 &
  \cellcolor[HTML]{B7E1CD}1 &
  \cellcolor[HTML]{B7E1CD}1 &
  \cellcolor[HTML]{B7E1CD}1 &
  \cellcolor[HTML]{72BE8D}\textbf{90.9} \\ \hline
P17 &
  \cellcolor[HTML]{B7E1CD}1 &
  \cellcolor[HTML]{B7E1CD}1 &
  \cellcolor[HTML]{B7E1CD}1 &
  \cellcolor[HTML]{B7E1CD}1 &
  \cellcolor[HTML]{F4CCCC}0 &
  \cellcolor[HTML]{B7E1CD}1 &
  \cellcolor[HTML]{B7E1CD}1 &
  \cellcolor[HTML]{B7E1CD}1 &
  \cellcolor[HTML]{B7E1CD}1 &
  \cellcolor[HTML]{B7E1CD}1 &
  \cellcolor[HTML]{B7E1CD}1 &
  \cellcolor[HTML]{72BE8D}\textbf{90.9} \\ \hline
P18 &
  \cellcolor[HTML]{B7E1CD}1 &
  \cellcolor[HTML]{B7E1CD}1 &
  \cellcolor[HTML]{B7E1CD}1 &
  \cellcolor[HTML]{B7E1CD}1 &
  \cellcolor[HTML]{B7E1CD}1 &
  \cellcolor[HTML]{B7E1CD}1 &
  \cellcolor[HTML]{B7E1CD}1 &
  \cellcolor[HTML]{B7E1CD}1 &
  \cellcolor[HTML]{B7E1CD}1 &
  \cellcolor[HTML]{B7E1CD}1 &
  \cellcolor[HTML]{B7E1CD}1 &
  \cellcolor[HTML]{57BB8A}\textbf{100} \\ \hline
P19 &
  \cellcolor[HTML]{B7E1CD}1 &
  \cellcolor[HTML]{B7E1CD}1 &
  \cellcolor[HTML]{B7E1CD}1 &
  \cellcolor[HTML]{B7E1CD}1 &
  \cellcolor[HTML]{B7E1CD}1 &
  \cellcolor[HTML]{B7E1CD}1 &
  \cellcolor[HTML]{B7E1CD}1 &
  \cellcolor[HTML]{B7E1CD}1 &
  \cellcolor[HTML]{B7E1CD}1 &
  \cellcolor[HTML]{B7E1CD}1 &
  \cellcolor[HTML]{B7E1CD}1 &
  \cellcolor[HTML]{57BB8A}\textbf{100} \\ \hline
P20 &
  \cellcolor[HTML]{B7E1CD}1 &
  \cellcolor[HTML]{B7E1CD}1 &
  \cellcolor[HTML]{B7E1CD}1 &
  \cellcolor[HTML]{B7E1CD}1 &
  \cellcolor[HTML]{B7E1CD}1 &
  \cellcolor[HTML]{B7E1CD}1 &
  \cellcolor[HTML]{B7E1CD}1 &
  \cellcolor[HTML]{B7E1CD}1 &
  \cellcolor[HTML]{B7E1CD}1 &
  \cellcolor[HTML]{B7E1CD}1 &
  \cellcolor[HTML]{B7E1CD}1 &
  \cellcolor[HTML]{57BB8A}\textbf{100} \\ \hline
\textbf{Feature \%} &
  \cellcolor[HTML]{57BB8A}\textbf{100} &
  \cellcolor[HTML]{57BB8A}\textbf{100} &
  \cellcolor[HTML]{FFF2CC}\textbf{35} &
  \cellcolor[HTML]{97B77D}\textbf{80} &
  \cellcolor[HTML]{A7B67A}\textbf{75} &
  \cellcolor[HTML]{97B77D}\textbf{80} &
  \cellcolor[HTML]{A7B67A}\textbf{75} &
  \cellcolor[HTML]{B7B577}\textbf{70} &
  \cellcolor[HTML]{A7B67A}\textbf{75} &
  \cellcolor[HTML]{A7B67A}\textbf{75} &
  \cellcolor[HTML]{D7B371}\textbf{60} &
  \multicolumn{1}{l|}{} \\ \hline
\end{tabular}%
}
\end{table}

Table \ref{tab:stats} shows the success rate of the think-aloud study by feature and teacher. Each row corresponds to a teacher, while each column represents a feature. Green boxes denote success, and red boxes indicate failure. The columns at the end (User Score) reflect the overall success percentage for each user. The final row (Feature \%) shows the aggregate success percentage for each feature. For additional details, refer to Appendix \ref{system}. Most teachers, i.e., 15 out of 20 (75\%), could navigate the system after two weeks, demonstrating its ease of use.  All teachers could access the chatbot to say hi to the chatbot (Bonjour), and they all completed the survey. Goal setting was the most complex feature, with only 35\% people being able to use it. Answering questions and accessing stories were moderately complex. Of the teachers, 15 out of 20 (75\%) used more than half of the features, and we categorized these users as \textit{experts}. We contrasted their usage patterns with the remaining \textit{novice} users.

\textit{Experts} (the top 80\%) exhibited nuanced usage of the \textit{chatbot}. For instance, they used the menu \textit{buttons} instead of typing the numbers, which improved their navigation speed. They also remembered that typing \textit{0} (default home) would return them to the \textit{home menu}, which helped them recover from complex navigation situations. Additionally, experts would zoom in on the relevant section of the menu to focus on relevant information instead of reading the entire text. They confidently navigated the current menu for subsequent tasks rather than restarting their search after every task. Another advantage observed among experts was using the chatbot in \textit{free basics mode}, which allowed teachers to access the chatbot without spending money for its data usage. Free basics from the MTN phone network allowed teachers to access Messenger and Facebook without incurring data charges. This was particularly beneficial for teachers in areas with limited access to paid data plans, as it encouraged greater engagement with the tool. Teachers could request features or seek help directly without incurring additional costs. 

However, while some teachers navigated these features successfully, \textit{novice} users struggled with reading, navigation, and providing input into the chatbot, suggesting a learning curve for some. \textit{Novices} (the remaining 30\%) often became overwhelmed by the tasks or the amount of information in the menus. They hesitated to attempt tasks, and the research assistant sometimes had to encourage them to proceed instead of giving up without trying. Novices frequently struggled with reading and comprehending the text in the menus, leading to slow navigation. They sometimes form incorrect mental models of the menu system (e.g., referencing the wrong menus or inputting incorrect menu options). Typing wrong inputs often led them to unintended menus, causing them to lose focus; the interviewer frequently had to remind them of their task. Additionally, novices often forgot how to return to the \textit{home menu}, making it difficult to complete or restart tasks.

These challenges highlight the need for additional support and training for new users. Teachers acknowledged that regular chatbot practice improved participation and helped them overcome initial usability challenges. Early on, many teachers struggled with the chatbot's features, as one participant shared, \textit{"In the early days when I had difficulties (with the chatbot), but over time I was able to access these features easily."} As teachers became more familiar with the tool, their confidence and ability to engage with its functionalities increased, leading to more effective usage. Another teacher remarked, \textit{"During the chatbot introductions, I had no idea about it, so it was not easy for me, but now I cannot say that I have mastered it, but it's okay."}

Lastly, we observed that all teachers typed very slowly, and often, each input took an average of 2--3 minutes to type approximately 30 words of text. Teachers did not use auto-suggest; instead, they typed every letter, reducing their speed. Additionally, some devices had physical limitations (e.g., scratches, cracks) or operated very slowly, making it difficult to type and use the applications effectively. The device limitations are prominent in prior work in ICTD discovered in basic phone usage by farmers in Kenya \cite{Wyche:doi:10.1080/02681102.2015.1048184}.

\section{Discussion}

While producing valuable results in the initial study, the Think-aloud method faced challenges when applied in a longitudinal setting in a subsequent study. During a longitudinal study later, we conducted Think-alouds four months after the chatbot deployment. We found it challenging to recruit enough intermediate or expert users who had engaged sufficiently with the chatbot to provide meaningful feedback on its usability. This lack of chatbot engagement among users made it challenging to gather representative samples and understand the usability challenges the chatbot presented. Importantly, this problem was not a limitation of the Think-aloud method itself but rather a reflection of insufficient usage by participants. Therefore, we reflected on the necessity for a minimum level of user interaction to apply the method in future studies. 

Another challenge in a subsequent study was inconsistent training among our research teams. Two teams were collecting data, and one team lead misunderstood the purpose of the think-aloud method, mistakenly intervening to assist participants when they encountered difficulties. The team's intervention contaminated the dataset and undermined the integrity of the think-aloud, rendering the collected data unusable. These experiences express the importance of consistent training for all team members to ensure the validity of the data and the effectiveness of the Think-aloud method.

Through our poster at ICTD, we expect to discuss our process of adapting the think-aloud session to an Ivorian context. The data collection team was new to HCI, and we had to adjust the think-aloud to their existing data collection practices, i.e., surveys and interviews. Thus, it was considerable work to explain the method. Additionally, the explanations had to be done remotely due to travel constraints. We also expect to discuss the data collected and its utility in helping us redesign the chatbot features. Lastly, we expect to discuss the opportunities for future research.


\begin{acks}
We thank our Pratham and TaRL Africa collaborators and the Ivorian Ministry for assisting in setting up the study. We are also very grateful to our field team members, Adji Yves, Fabrice Tanoh, Hermann Apke, Salem Konan, and the entire Proterrain team, for their contributions to our research. Lastly, we acknowledge the financial support from the Jacobs Foundation.
\end{acks}

\bibliographystyle{ACM-Reference-Format}
\bibliography{sample-base,references,bib-refs}

\appendix

\section{System Design}\label{system}

The flow chart depicts the various features of the \chatbot and the menu hierarchy for accessing them. We designed the \chatbot using USSD menus(menus with numeric options), which are interactions familiar to the context. Users would be presented with a menu with numerical options, leading to subsequent menus or features. 

\begin{figure}[!htp]
\centering
  \includegraphics[width=0.95\columnwidth]{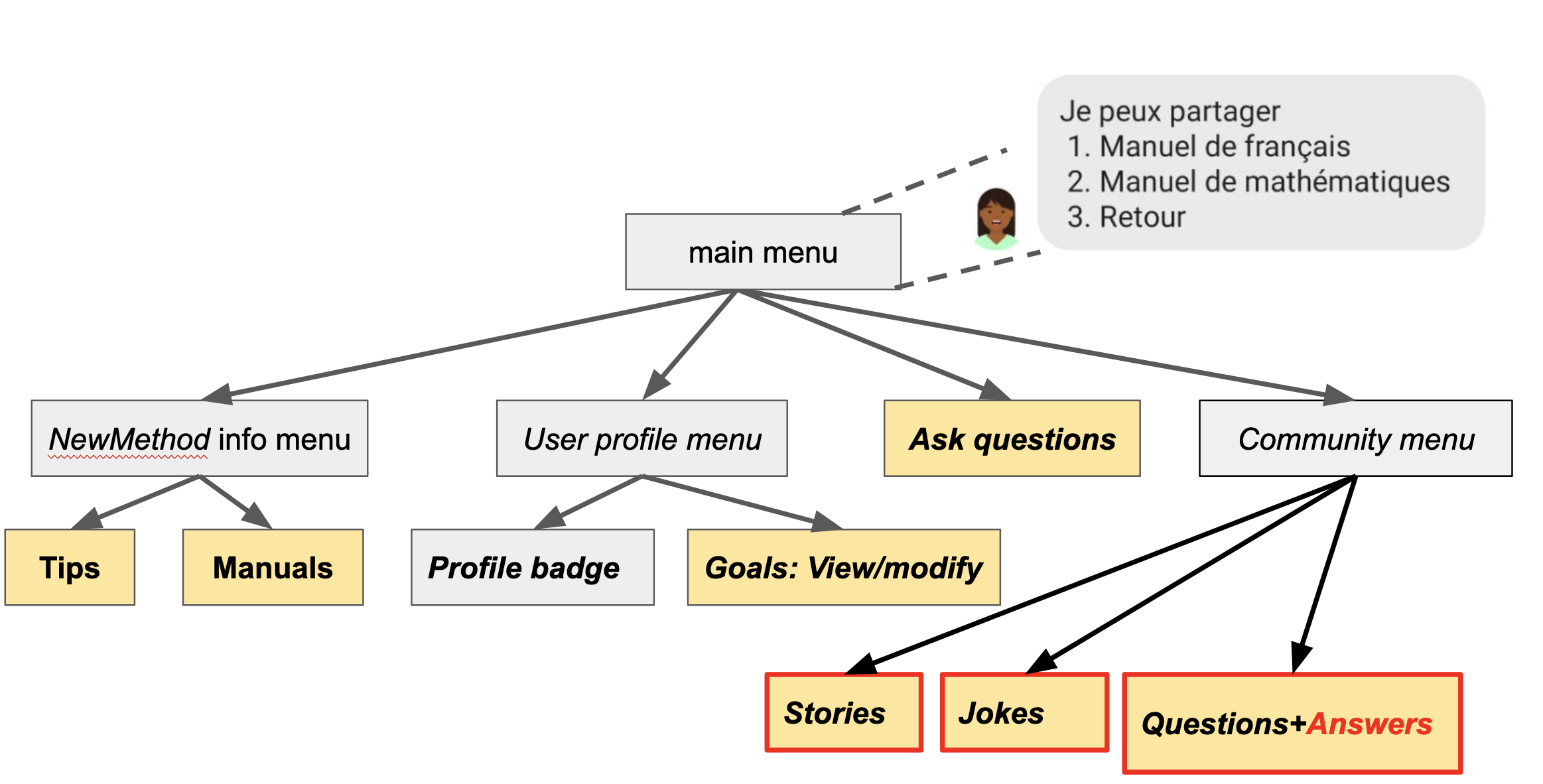}
  \caption{Flowchart of system features used in the study   }~\label{fig:system4}
\end{figure}

These features were tested through the think-aloud sessions, shown in Table \ref{tab:stats}.
\begin{itemize}

\item {\verb|Main Menu (Bonjour)|}: The main menu was the default content shown to the users. Users could always access it by typing "0". 

\item {\verb|Introductory survey|}: There were introductory survey questions with example interactions to onboard the teachers into the system. These questions were intended as \chatbot training sessions during the teacher training. 

\item {\verb|Goals|}: The profile menu gave users access to the Goals feature. The goals option allowed users to view and modify previously selected goals. These were supposed to support the teachers reflect on their work for the week. 

\item {\verb|Browser|}: Teachers were asked to browse the different menu options. 

\item {\verb|Tips|}: \TaRL Information menu led users to the Tips options. Tips showed curated short content about the \TaRL program from the official documents. 

\item {\verb|Manuals|}: \TaRL manuals gave users access to offical documents related to \TaRL for French and Math content. 

\item {\verb|Questions|}: Users could directly interact with the \chatbot by asking questions. However, we added a menu option to guide users who may need additional scaffolding. We had the human-AI hybrid interaction with data from the \TaRL manual.

\item {\verb|Community Menu|}: We also had the community feature where users could (a) share and read experiences about \TaRL,  (b) share and read jokes, or (c) answer curated questions. These were experimental, with sample content curated by the researchers.

\end{itemize}
\end{document}